\documentclass[12pt]{iopart}
\usepackage{hyperref}
\usepackage{epsfig}
\usepackage[numbers]{natbib}

\usepackage{setstack}
\usepackage{iopams}
\usepackage[latin1]{inputenc}
\usepackage{url}

 \providecommand{\email}[1]{\href{mailto:#1}{\texttt{#1}}}
\providecommand{\dprod}{\! \cdot \!}%

\begin{document}
%

\title[An hypersphere model of the Universe]{An hypersphere model of the Universe
-- The dismissal of dark matter}
\author{Jos\'e B. Almeida}
\address{Universidade do Minho, Departamento de F\'isica,
4710-057 Braga, Portugal.}

\ead{\email{bda@fisica.uminho.pt}}


\date{\today}

\begin{abstract}                
One can make the very simple hypothesis that the Universe is the inside of an
hypersphere in 4 dimensions, where our 3-dimensional world consists of
hypersurfaces at different radii. Based on this assumption it is possible to
show that Universe expansion at a rate corresponding to flat comes as a direct
geometrical consequence without intervening critical density; any mass density
is responsible for opening the Universe and introduces a cosmological constant.
Another consequence is the appearance of inertia swirls of expanding matter,
which can explain observed velocities around galaxies, again without the
intervention of dark matter. When restricted to more everyday situations the
model degenerates in what has been called 4-dimensional optics; in the paper
this is shown to be equivalent to general relativity in all static isotropic
metric situations. In the conclusion some considerations bring the discussion
to the realm of 4D wave optics.
\end{abstract}

\submitto{\JPA}
\maketitle

\section{Introduction}
In this work I make the simple hypothesis that the Universe can be modelled as
the volume of an hypersphere in 4 Euclidean dimensions. Naturally the position
vector for any point has one single coordinate, the distance to the center of
the hypersphere, but displacements have all 4 coordinates: one distance and 3
angles. It is easy to evaluate the length of any displacement and one can
easily conclude that for small displacements, provided the distance to the
center is large, the angles can be replaced by distances on a plane tangent to
an hyperspherical surface, providing a local Euclidean frame for the study of
displacements.

In general, though, the hyperspherical nature of the space has important
consequences. It is shown that by assigning the meaning of time to the length
of displacements one concludes that the distance between any two points in a
3-dimensional hypersurface  space increases at a rate proportional to the
distance; this is exactly what one finds in our Universe but is derived as a
consequence of geometry and not of any critical mass density. A similar
argument applied to rotary motion allows the conclusion that this is a natural
form of inertial movement and can be applied to galaxies' dynamics to explain
the exceedingly large orbital velocities that are detected. Here too, geometry
and not hidden mass is the main cause of movement. Naturally mass densities are
important for the detailed analysis of observations but they are responsible
only for perturbations of a global phenomenon with geometrical causes; it is
shown that any mass density is responsible for opening the Universe as well as
for a cosmological constant.

On a small scale the space becomes nearly Euclidean and it must be shown that
this space is adequate for the description of classical mechanics, at least as
effectively as general relativity does; dynamics in Euclidean 4-space is called
\emph{4-dimensional optics (4DO)} because it is governed by an extension of
Fermat's principle. The paper demonstrates full equivalence between dynamics in
hyperbolic general relativity space and 4DO for the case of static isotropic
metrics; the particular case of Schwarzschild's metric is analyzed and an
exponential metric offering the same predictions as Schwarzschild's is
proposed.

\section{4-dimensional hyperspheric coordinates}
As an introduction to 4-dimensional hyperspheric coordinates it is useful to
revise the case of spherical coordinates in 3 dimensions. The position vector
for any point is always written $s = r \sigma_r$, where $\sigma_r$ is a unitary
vector. If needed we can always express $\sigma_r$ in terms of the orthonormed
frame $\{\sigma_1, \sigma_2, \sigma_3\}$
\begin{equation}
    \label{eq:sigmar}
    \sigma_r = \sin \theta \cos \phi\, \sigma_1 + \sin \theta \sin \phi\, \sigma_2
    + \cos \theta\, \sigma_3.
\end{equation}
We say that $\{\sigma_1, \sigma_2, \sigma_3\}$ is a fiducial frame because it
is orthonormed and its vectors don't rotate in a displacement.

A displacement in spherical coordinates is the vector
\begin{equation}
    \label{eq:dsspherical}
    \mathrm{d}s = \partial_r s\, \mathrm{d}r + \partial_\theta s\, \mathrm{d} \theta +
    \partial_\phi s\, \mathrm{d} \phi,
\end{equation}
with $\partial_\mu$ representing partial derivative with respect to coordinate
$\mu$. Resorting to the fiducial frame we can establish the derivatives of
$\sigma_r$
\begin{equation}
    \label{eq:sigmarderiv}
    \partial_r \sigma_r = 0,~~~~ \partial_\theta \sigma_r = \sigma_\theta, ~~~~
    \partial_\phi \sigma_r = \cos \theta\, \sigma_\phi,
\end{equation}
where $\{\sigma_r, \sigma_\theta, \sigma_\phi\}$ form a new orthonormed frame
which is not a fiducial frame because its vectors rotate.
\begin{eqnarray}
    \sigma_\theta &= \cos \theta \cos \phi\, \sigma_1 + \cos \theta \sin \phi \,
    \sigma_2 - \sin \theta \sigma_3, \\
     \sigma_\phi &= -\sin \phi\, \sigma_1 +
    \cos \phi\, \sigma_2.
\end{eqnarray}
We can express this rotation by a set of partial derivatives
\begin{eqnarray}
    \partial_r \sigma_r = 0, ~~~~ &\partial_\theta \sigma_r = \sigma_\theta,
     ~~~~ &\partial_\phi \sigma_r = \sin \theta \sigma_\phi, \nonumber \\
    \partial_r \sigma_\theta = 0, ~~~~ &\partial_\theta \sigma_\theta =
    -\sigma_r, ~~~~ &\partial_\phi \sigma_\theta = \cos \theta \sigma_\phi, \\
    \partial_r \sigma_\phi = 0, ~~~~ &\partial_\theta \sigma_\phi = 0, ~~~~
    &\partial_\phi \sigma_\phi = -\sin \theta\, \sigma_r - \cos \theta\, \sigma_\theta.
    \nonumber
\end{eqnarray}

The displacement vector can now be found by application of the derivatives to
Eq.\ (\ref{eq:dsspherical})
\begin{equation}
    \label{eq:dsspherical2}
    \mathrm{d}s = \sigma_r \mathrm{d}r + r\sigma_\theta \mathrm{d} \theta + r
    \sin \theta \sigma_\phi \mathrm{d}\phi.
\end{equation}
A coordinate frame for spherical coordinates can't be $\{\sigma_r,
\sigma_\theta, \sigma_\phi\}$, however, because the general definition for a
coordinate frame is $g_\mu = \partial_\mu s$, \cite{Hestenes86:2}. Using Eq.\
(\ref{eq:dsspherical2}) we can write
\begin{equation}
    \label{eq:spherframe}
    g_r = \sigma_r,~~~~ g_\theta = r \sigma_\theta, ~~~~ g_\phi = r \sin
    \theta\, \sigma_\phi.
\end{equation}

The displacement vector $\mathrm{d}s$ is now written in the general form
\begin{equation}
    \mathrm{d}s = g_j \mathrm{d}x^j;
\end{equation}
where the index $j$ is replaced by $(r, \theta, \phi)$, $x^\mu$ represents the
coordinates $r, \theta, \phi$, respectively, and the summation convention for
repeated indices is used. Defining the metric tensor elements $g_{j j} = g_j
\dprod g_j$ we can evaluate an interval by
\begin{equation}
    (\mathrm{d} s)^2 = \mathrm{d}s \dprod \mathrm{d}s = g_{\mu
    \nu} \mathrm{d}x^j \mathrm{d}x^j.
\end{equation}

The spherical coordinates example can now be easily extended to a general
situation in 4 dimensions.
We will consider 4-dimensional space with hyperspheric symmetry where $R$ is
the distance to the origin and $\alpha^j$, $j = 1,2,3$ are angles. The position
vector is naturally $s = R \sigma_0$, with $\sigma_0$ the unit vector of the
radial direction; the displacement vector is obtained by extrapolation of Eq.\
(\ref{eq:dsspherical2})
\begin{equation}
    \label{eq:ds4dsphere}
    \mathrm{d}s = \mathrm{d}R \sigma_0 + R \left(\mathrm{d}\alpha^1
    \sigma_1 +
    \sin \alpha^1 \mathrm{d} \alpha^2 \sigma_2 + \sin \alpha^1 \sin \alpha^2
    \mathrm{d} \alpha^3
    \sigma_3 \right).
\end{equation}
If the displacements are small compared to the hypersphere radius $R$, we can
choose a privileged origin for the angles such that all the angles are small
and the sines become unity.
\begin{equation}
    \label{eq:ds4do}
    \mathrm{d}s = \mathrm{d}R \sigma_0 +
     R \left(\mathrm{d}\alpha^j \sigma_j\right).
\end{equation}
We will now define the new coordinates $x^j = R \alpha^j$ so that
$\mathrm{d}x^j = \mathrm{d}R \alpha^j + R \mathrm{d}\alpha^j$. Inverting the
relation it is $R \mathrm{d}\alpha^j = \mathrm{d}x^j - \mathrm{d}R x^j/R$.
Replacing above
\begin{equation}
    \label{eq:ds4do2}
    \mathrm{d}s = \mathrm{d}R \sigma_0 +
    \left(\mathrm{d}x^j- \frac{\mathrm{d}R}{R}\, x^j\right)\sigma_j.
\end{equation}
And the displacement length is evaluated by
\begin{equation}
    \label{eq:4dointerval}
    (\mathrm{d}s)^2 = (\mathrm{d}R)^2  +
    \sum \left(\mathrm{d}x^j- \frac{\mathrm{d}R}{R}\, x^j\right)^2.
\end{equation}
There is no reason why the displacement should not be given in time units, as
long as we use some length and time standards, $\mathcal{L}$ and $\mathcal{T}$
respectively, and replace $\mathrm{d}s = \mathrm{d}t\mathcal{L}  /\mathcal{T}$;
as a consequence $\mathcal{L/T} =c$ is the speed of light in vacuum.
\begin{equation}
    \label{eq:dts4do}
    (\mathrm{d}t)^2 =\left(\frac{\mathcal{T}}{\mathcal{L}}\right)^2
    \left[(\mathrm{d}R)^2  +
    \sum \left(\mathrm{d}x^j- \frac{\mathrm{d}R}{R}\, x^j\right)^2 \right].
\end{equation}
Dividing both members by $(\mathrm{d}t)^2$
\begin{equation}
    \label{eq:empty}
    1 =\left(\frac{\mathcal{T}}{\mathcal{L}}\right)^2
    \left[(\dot{R})^2  +
    \sum \left(\dot{x}^j- \frac{\dot{R}}{R}\, x^j\right)^2 \right]
\end{equation}

We are going to interpret the coordinate $R$ as the time elapsed from the
Universe's origin, albeit measured as length, and coordinates $x^j$ as being
the usual $x,y,z$ coordinates of 3-dimensional space. We will develop the
consequences of this interpretation in the following paragraphs.
\section{Free space dynamics}
Examining displacements on 3D hypersurface we make $\dot{R} = 0$ in Eq.\
(\ref{eq:empty})
\begin{equation}
    \label{eq:lightspeed}
    1 = \left(\frac{\mathcal{T}}{\mathcal{L}}\right)^2
     \sum (\dot{x}^j)^2 = \left(\frac{\mathcal{T}}{\mathcal{L}}\,c \right)^2.
\end{equation}
Light travels with velocity $c$ in 3-space and the model can accommodate it by
zeroing the displacement in the radial direction; photons follow great circles
of constant $R$.

Proceeding to the analysis of massive particle's dynamics we note that the
Euler-Lagrange equations for the geodesics of any Riemanian space can be
derived from a constant Lagrangian, made equal to $1/2$ for convenience
\cite{Martin88}. Using Eq.\ (\ref{eq:ds4do}) we can evaluate $\mathrm{d}s^2 =
\mathrm{d}s \dprod \mathrm{d}s$ and divide both members by $\mathrm{d}s^2$; the
first member is then made equal to twice the Lagrangian
\begin{equation}
    \label{eq:alphalag}
    1 = 2L = \dot{R}^2 + R^2 \sum (\dot{\alpha}^j)^2;
\end{equation}
the four conjugate momenta are
\begin{eqnarray}
    \label{eq:pj}
    p_j &= R^2 \dot{\alpha}^j =A^j \\
    p_0 &= \dot{R}.
\end{eqnarray}
The four $A^j$ are conserved quantities because the Lagrangian is independent
from $\alpha_j$; $A^j \sigma_j$ is a vector whose norm is $A$. Replacing in
Eq.\ (\ref{eq:alphalag})
\begin{equation}
    1 = \dot{R}^2 + \left(\frac{A}{R}\right)^2.
\end{equation}
Upon integration, with appropriate choice for the origin of time, we obtain the
solution
\begin{eqnarray}
    \label{eq:rdet}
    R = \sqrt{A^2 + t^2},  \\
    \dot{R} = \frac{t}{\sqrt{A^2 + t^2}}.
\end{eqnarray}

Returning to linear rather than angular coordinates and considering Eq.\
(\ref{eq:pj})
\begin{equation}
    \dot{x}^j = R \dot{\alpha}^j + \frac{\dot{R}}{R}\, x^j.
\end{equation}
Inserting Eqs.\ (\ref{eq:pj}) and (\ref{eq:rdet})
\begin{equation}
    \label{eq:Hubble1}
    \frac{\dot{x}^j}{x^j} = \frac{A^j}{x^j} + \frac{t}{A^2 + t^2} ;
\end{equation}
the first member defines the Hubble parameter $H$ and the second member tells
us that the velocity does not stay constant but approximately with $t x^j
/A^2$.

Analyzing Eq.\ (\ref{eq:empty}) we have to decide if and when the term $\dot{R}
x^j /R$ can be neglected in face of $\dot{x}^j$. The condition we want can be
expressed by
\begin{equation}
    \frac{x^j}{\dot{x}^j}\, \ll \frac{R}{\dot{R}}\, ;
\end{equation}
we have a comparison between two times: on the first member the time it would
take a distant body to travel to the origin of the laboratory coordinates and
on the second member another time which we will assign below to the tame it
takes light to travel from the confines of the Universe. This condition is met
for nearby objects which are not moving exceedingly slow; when it can be met
Eq.\ (\ref{eq:empty}) reduces to
\begin{equation}
    \label{eq:maxspeed}
    (\dot{R})^2 + \sum (\dot{x}^j)^2 = c^2,
\end{equation}
placing an upper limit on the speed of moving particles. It is also apparent
that the movement of masses implies that they move outwards in the hypersphere
through $\dot{R}$.

Returning to Eq.\ (\ref{eq:alphalag}) it is easy to conclude that for bodies
comoving with the Universe's expansion we must have constant $\alpha^j$ and
$\dot{R} =c$, so the Universe must be expanding at the speed of light. For the
distance coordinates we get
\begin{equation}
    \dot{\alpha}^j = \frac{\dot{x}^j R - \dot{R}x^j}{R^2} = 0.
\end{equation}
According to the above argument
\begin{equation}
    \label{eq:Hubble}
    \frac{\dot{x}^j}{x^j} = \frac{\dot{R}}{R} = \frac{c}{R} = H;
\end{equation}
H is the Hubble parameter and its measurement gives us the size of present day
Universe. If we use for the Hubble parameter a value of $81~ \mathrm{km}~
\mathrm{s}^{-1}/~ \mathrm{Mpc}$ the resulting size for the Universe is $1.2
\times 10^{10}~ \mathrm{ly}$. Additionally, considering the Universe's
expansion is influenced by its mass density so that $A$ is effectively
positive, Eq.\ (\ref{eq:Hubble1}) tells us the effective Hubble parameter will
also increase with time, approximately with $t/A^2$.

The constant orbital velocity observed in the periphery of most galaxies
($\omega r =$ constant) is one of the big puzzles in the Universe which is
normally explained with recourse to massive halos of dark matter
\cite{Silk97,Narlikar02}, although some have tried different explanations with
limited success; for instance \citet{Milgrom83, Milgrom83a} modified Newton
dynamics empirically. Below we look at the predictions of the hypersphere model
for orbital velocities to verify that such explanations are not needed if one
accepts that the universe is expanding as an hypersphere.

The gravitational field on the periphery of a galaxy must be negligible without
the dark matter halo contribution. The question we will try to answer is
whether the Universe expansion can drive a rotation, once the material has been
set in motion by some other means. In the affirmative case we must find out if
the rotation speed can be kept invariant with distance to the center, as
observed in galaxies. Recalling Eq.\ (\ref{eq:ds4do2}) we will rewrite this
equation in spherical coordinates
\begin{equation}
    \label{eq:dsradial}
    ds = \mathrm{d}R \sigma_0 + \mathrm{d}r \sigma_r + r\mathrm{d} \theta
    \sigma_\theta + r \sin \theta \mathrm{d}\phi \sigma \phi - \frac{d R}{R}\, r
    \sigma_r.
\end{equation}
Notice the last term and compare it to Eq.\ (\ref{eq:ds4do2}); we have replaced
$x^j \sigma_j$ by $r \sigma_r$ in a standard passage from Cartesian to
spherical coordinates. It is usual to make $\theta = \pi/2$ whenever dealing
with orbits, because we know in advance that orbits are flat. Defining
$\mathrm{d}t^2 = \mathrm{d}s^2$ and calling $v$ to $\mathrm{d} s/\mathrm{d}t$
we can write
\begin{equation}
    v = \dot{R} \sigma_0 + \left(\dot{r} - \frac{\dot{R} r}{R} \right) \sigma_r
    + r \dot{\phi} \sigma_\phi.
\end{equation}
If the parenthesis vanishes the movement becomes circular without any central
potential; it is driven solely by the galaxy expanding at the same rate as the
Universe. The equation above shows that $r \dot{\phi}= $ constant is the
natural inertia condition for the hyperspheric Universe; swirls will be
maintained by a radial expansion rate which exactly matches the quotient
$\dot{R}/R$. In any practical situation $\dot{R}$ will be very near the speed
of light and the quotient will be virtually equal to the hubble parameter; thus
the expansion rate for sustained rotation is $\dot{r}/r =H$. If applied to our
neighbor galaxy Andromeda, with a radial extent of $30~\mathrm{kpc}$, using the
Hubble parameter value of $81~\mathrm{km}~\mathrm{ s}^{-1}/\mathrm{Mpc}$, as
above, the expansion velocity is about $2.43~\mathrm{km}~\mathrm{ s}^{-1}$;
this is to be compared with the orbital velocity of near $300~\mathrm{km}~
\mathrm{s}^{-1}$.

The model proposed for galaxy dynamics consists of a core dominated by
gravitational and electromagnetic interactions from which some material escapes
and starts swirling by inertia, while continuing to be accelerated by the
remnants of gravity; near the periphery all the gravity is extinct and only
inertial rotations prevails.

\section{Curved space dynamics}
The hypersphere model would be useless if it could not be made compatible with
classical mechanics in everyday situations; in this paragraph we will see that
full compatibility exists.

Equation (\ref{eq:dts4do}) with the constraint $x^j\ll R$ defines 4D Euclidean
space, with signature $(++++)$, which differs from Minkowski spacetime with
signature $(+---)$. If we use $x^0$ to represent $R$ the interval of that space
is given by
\begin{equation}
    \label{eq:isointerval}
    (\mathrm{d}t)^2 = \frac{1}{c^2}\, \sum_\mu (\mathrm{d}x^\mu)^2.
\end{equation}
In this space Eq.\ (\ref{eq:maxspeed}) establishes that everything moves with
the speed of light and it becomes natural to extend to 4-space Fermat's
principle which governs geometric optics in 3D
\begin{equation}
    \label{eq:Fermat}
    \delta \int^{P_2}_{P_1}n \mathrm{d}s =0,
\end{equation}
where $n$ is a function of coordinates $1$ to $3$, called \emph{refractive
index}, defined as the ratio between local 4-speed and the speed of light in
vacuum.
\begin{equation}
    n = \frac{1}{v}=\frac{\mathrm{d}t}{\mathrm{d}s}.
\end{equation}
The extension of Fermat's principle to 4D justifies our use of the designation
\emph{4-dimensional optics} to refer the study of 4D dynamics and  wave
propagation; we will use the acronym \emph{4DO} as a substitute for the full
designation. From this point onwards we will make $c=1$ following the uses of
general relativity papers, which corresponds to using actual displacements
measured in length rather than time units.

In an homogeneous medium Eq.\ (\ref{eq:Fermat}) states that trajectories are
straight lines in 4-space; in particular when $n=1$, everything moves with
4-velocity with modulus equal to the speed of light in vacuum. Geometric optics
in 3D becomes a direct consequence of 4DO and is obtained from Eq.\
(\ref{eq:Fermat}) by setting $\mathrm{d}x^0=0$, in agreement with our previous
contention that photons travel on 3D space;
\begin{equation}
    \delta \int^{R_2}_{R_1}n \mathrm{d}l =0,
\end{equation}
with $(\mathrm{d}l)^2 =\sum (\mathrm{d}x^j)^2$ and $j = 1 \ldots3$.

The integrand in Fermat's principle, $n\mathrm{d}s$, can be replaced by
$\mathrm{d}t$, allowing its interpretation with the phrase: \emph{Radiation and
massive bodies travel between two points in 4-space along a path which makes
time interval an extremum.} Using $(\mathrm{d}s)^2$ from Eq.\
(\ref{eq:isointerval}) the time interval is given by
\begin{equation}
    (\mathrm{d}t)^2={n^2}\sum
    (\mathrm{d}x^\mu)^2.
\end{equation}
This can be generalized further without considering non-isotropic media;
\begin{equation}
    \label{eq:isotropic}
    (\mathrm{d}t)^2=(n_0 \mathrm{d}x^0)^2+
    {(n_r)^2}\sum
    (\mathrm{d}x^i)^2.
\end{equation}
The anisotropy relative to coordinate $x^0$ is not apparent in 3 dimensions,
and the medium can still be classified as isotropic.
An alternative interpretation of Eq\. (\ref{eq:isotropic}) is in terms of
interval of curved isotropic space; it is equivalent saying that particles and
radiation travel slower than in vacuum in a given region of space and saying
that in the same region space is curved. Following the standard Lagrangian
choice
\begin{equation}
    \label{eq:isolag}
    1= 2  L =(n_0 \dot{x}^0)^2+
    {(n_r)^2}\sum
    (\dot{x}^i)^2.
\end{equation}
The Lagrangian is independent from $x^0$, so we have a conservation equation
\begin{equation}
    \label{eq:dotx0}
    (n_0)^2 \dot{x}^0 = \frac{1}{\gamma}.
\end{equation}
Replacing above,
\begin{equation}
    \label{eq:path}
    1= \frac{1}{(n_0)^2 \gamma^2}+
    {(n_r)^2}\sum
    (\dot{x}^i)^2.
\end{equation}
The remaining 3 Euler-Lagrange equations for the trajectory can be written
\begin{equation}
    \frac{\mathrm{d}}{\mathrm{d}t}\left( \frac{\partial
    L}{\partial \dot{x}^i}\right) = \partial_i L;
\end{equation}
replacing,
\begin{equation}
    \frac{\mathrm{d}}{\mathrm{d}t}\left[(n_r)^2 \dot{x}^i\right] =
    n_0 \partial_i n_0 (\dot{x}^0)^2
     +n_r \partial_i n_r \sum
    (\dot{x}^j)^2.
\end{equation}
Expanding the 1st member, inserting Eq.\ (\ref{eq:dotx0}) and rearranging
\begin{equation}
    \ddot{x}^i = \frac{n_0 \partial_i n_0 }{n_r^2}\,(\dot{x}^0)^2 -
    \frac{\partial_i n_r}{n_r} \sum
    (\dot{x}^j)^2.
\end{equation}

The previous equation must now be compared to the predictions of general
relativity. A general static relativistic interval for isotropic coordinates
can be written
\begin{equation}
    (\mathrm{d}s)^2 = \left(\frac{1}{n_0}\right)^2
    (\mathrm{d}t)^2 - \left(
    \frac{n_r}{n_0} \right)^2 \sum (\mathrm{d} x^i)^2.
\end{equation}
Since $n_0$ and $n_r$ are arbitrary functions of coordinates $x^j$, this form
allows all possibilities. A suitable Lagrangian for this space's geodesics is
\begin{equation}
    2L = 1 = \left(\frac{1}{n_0}\right)^2\left(\frac{\mathrm{d}t}
    {\mathrm{d}s}\right)^2 - \left(
    \frac{n_r}{n_0} \right)^2 \sum \left(\frac{\mathrm{d} x^i}
    {\mathrm{d}s}\right)^2.
\end{equation}
There is a conserved quantity because the Lagrangian does not depend on $t$
\begin{equation}
    \frac{1}{(n_0)^2}\,\frac{\mathrm{d}t}{\mathrm{d}s}
    = \gamma .
\end{equation}
Replacing in the lagrangian $\mathrm{d}/\mathrm{d}s \rightarrow
\mathrm{d}/\mathrm{d}t \times \mathrm{d}t/\mathrm{d}s$ we obtain again Eq.\
(\ref{eq:path}):
\begin{equation}
    \frac{1}{(n_0)^2 \gamma^2}
    =1 - (n_r)^2 \sum (\dot{x}^i)^2.
\end{equation}

We conclude that at least for static isotropic metrics the geodesics of general
relativity can be mapped to those of 4DO and so it is a matter of personal
preference which formalism each one uses. We believe that the proof can be
extended to all static metrics but that is immaterial for the present work.

We will now look at Schwarzschild's metric to see how it can be transposed to
4D optics. We will have to use the dimensionless variable $G m/(c^2 r)$, where
$G$ is the gravitational constant. Since  a dimensionless variable can be built
with $\mathcal{L} m/(\mathcal{M}r)$, where $\mathcal{M}$ is the mass standard,
we will choose $\mathcal{M}= G \mathcal{L}/c^2=  G \mathcal{T}^2/\mathcal{L}$
and avoid constants in the expressions.

The usual form of Schwarzschild's metric is
\begin{equation}
    \mathrm{d}s^2 = \left(1-\frac{2m}{\rho} \right)\mathrm{d}t^2
    -\left(1-\frac{2m}{\rho} \right)^{-1}\mathrm{d}\rho^2 - \rho^2
    (\mathrm{d}\theta^2 + \sin^2 \theta \mathrm{d}\phi^2 ).
\end{equation}
This form is non-isotropic but a change of coordinates can be made that returns
an isotropic form  \citet[section 14.7]{Inverno96}:

\begin{equation}
    r=\left(\rho-m+\sqrt{\rho^2-2m \rho}\right)/2;
\end{equation}
and the new form of the metric is
\begin{equation}
    \mathrm{d}s^2 =
    \left(\frac{1-\frac{m}{2r}}{1+\frac{m}{2r}}\right)^2
    \mathrm{d}t^2 -
    \left(1+ \frac{m}{2r}\right)^4\left[ \mathrm{d}r^2 - r^2 (\mathrm{d}\theta^2
    + \sin^2 \theta \mathrm{d}\phi^2 ) \right].
\end{equation}
This corresponds to the refractive indices
\begin{equation}
    n_0 = \frac{1+ \frac{m}{2r}}{1-\frac{m}{2r}},
    ~~~~ n_r = \frac{(1 + \frac{m}{2r})^3}{1-\frac{m}{2r}},
\end{equation}
which can then be used by 4DO in Euclidean space.

We turn now to the constraints on the refractive indices so that experimental
data on light bending and perihelium advance in closed orbits can be predicted.
Light rays are characterized by $\mathrm{d}x^0=0$ in 4DO or by $\mathrm{d} s
=0$ in general relativity; the effective refractive index for light is then
\begin{equation}
    \sqrt{\frac{1}{\sum (\dot{x}^i)^2}} = n_r.
\end{equation}
For compatibility with experimental observations $n_r$ must be expanded in
series as (see \cite{Will01})
\begin{equation}
    n_r = 1 +\frac{2 m}{r} + O(1/r)^2.
\end{equation}
This is the bending predicted by Schwarzschild's metric and has been confirmed
by observations.

For the analysis of orbits its best to rewrite Eq.\ (\ref{eq:isolag}) for
spherical coordinates; since we know that orbits are flat we can make $\theta =
\pi/2$
\begin{equation}
    n_0^2 \dot{\tau}^2 + n_r^2 (\dot{r}^2 + r^2 \dot{\phi}^2 ) =1.
\end{equation}
The metric depends only on $r$ and we get two conservation equations
\begin{equation}
    n_0^2 \dot{\tau} = \frac{1}{\gamma},~~~~ n_r^2 r^2 \dot{\phi}
    = J.
\end{equation}
Replacing
\begin{equation}
    \frac{1}{\gamma^2 n_0^2}\, + n_r^2 \dot{r}^2 +
    \frac{J^2}{n_r^2 r^2}\, = 1.
\end{equation}
The solution of this equation calls for a change of variable $r = 1/u$; as a
result it is also $\dot{r} =\dot{\phi} \mathrm{d}r/\mathrm{d}\phi$; replacing
in the equation and rearranging
\begin{equation}
    \left(\frac{\mathrm{d}u}{\mathrm{d}\phi} \right)^2 =
    \frac{n_r^2}{J^2}\, - \frac{n_r^2}{J^2 \gamma^2 n_0^2}\, -u^2.
\end{equation}

To account for light bending we know that $n_r \approx 1+2m u$. For $n_0$ we
need 2nd order approximation \cite{Will01}, so we make $n_0 \approx 1 + \alpha
m u + \beta m^2 u^2$. We can also assume that velocities are low, so $\gamma
\approx 1$
\begin{equation}
    \left(\frac{\mathrm{d}u}{\mathrm{d}\phi} \right)^2 \approx
     \frac{2
    \alpha m}{J^2}\, u + \left(-1 + \frac{8 \alpha m^2}{J^2} -
    \frac{3 \alpha^2 m^2}{J^2} + \frac{2 \beta m^2}{J^2} \right)
    u^2.
\end{equation}
For compatibility with Kepler's 1st order predictions $\alpha =1$; then, for
compatibility with observed planet orbits, $\beta = 1/2$. Together with the
constraint for $n_0$, these are the conditions that must be verified by the
refractive indices to be in agreement with experimental data.

We know, of course, that the refractive indices corresponding to
Schwarzschild's metric verify the constraints above, however that is not the
only possibility. Schwarzschild's metric is a consequence of Einstein's
equations when one postulates that vacuum is empty of mass and energy, but the
same does not necessarily apply in 4DO. Leaving an open question about what
equations should be the counterparts of Einstein's in 4DO, one interesting
possibility for the refractive indices, in full agreement with observations, is
provided by
\begin{equation}
    \label{eq:n0}
    n_0 = \mathrm{e}^{m /r} \approx 1 + \frac{m}{r}\, + \frac{m^2}{2r^2}\,,
\end{equation}
\begin{equation}
    \label{eq:nr}
    n_r = \mathrm{e}^{2 m/r} \approx 1 + \frac{2 m}{r}\,.
\end{equation}

These refractive indices are as effective as those derived from Schwarzschild's
metric for light bending and perihelium advance prediction, although they do
not predict black holes. There is a singularity for $r=0$ which is not a
physical difficulty since before that stage quantum phenomena have to be
considered and the metric ceases to be applicable; in other words, we must
change from geometric to wave optics approach.

We shall look now at how the overall mass density in the Universe affects its
expansion rate; for this we will adopt the refractive indices from Eqs:\
(\ref{eq:n0}) and (\ref{eq:nr}) and we will denote $n_0 = n$ and $n_r=n^2$. The
radial equation Eq.\ (\ref{eq:dsradial}) can be used to construct a geodesic
equation modified with the introduction of $n$; we are only interested in
radial trajectories so $\mathrm{d}\theta = \mathrm{d}\phi=0$;
\begin{equation}
    c^2 = n^2 \dot{R}^2 + n^4 \dot{r}^2 + n^4 \left(\frac{\dot{R}}
    {R} \right)^2 r^2.
\end{equation}
Rearranging and noting that $H = \dot{r}/{r}$ is the measured Hubble parameter
\begin{equation}
    H^2 = \left(\frac{\dot{R}}{R} \right)^2 + \left(\frac{c^2}{n^4}-
    \frac{\dot{R}^2}{n^2} \right)\frac{1}{r^2}
\end{equation}

Looking at the definition for $n$, Eq.\ (\ref{eq:n0}), we note that $m$ is the
mass internal to a sphere of radius $r$ and must remain constant $m = 4 \pi
\rho_0 r_0^3$, where $\rho_0$ is the present density and $r_0$ is the present
radius of the sphere. Approximating the exponentials to the first order terms
\begin{equation}
    H^2 \approx \left(\frac{\dot{R}}{R} \right)^2 +\frac{c^2 - \dot{R}^2}{r^2}
    +\frac{(2\dot{R}^2 - 4 c^2)m}{r^3}.
\end{equation}
This equation is similar to Friedman's and can be interpreted as follows: when
the mass density is zero the exponentials are unity, we have seen that $\dot{R}
=c$ and the Hubble parameter is $H = c/R$; in the original Friedman equation
this situation corresponds to a flat Universe and is attributed to a critical
density, whose source is attributed to dark matter. Any mass density will make
$\dot{R}<c$ and the second term produces an open Universe; the 3rd term is
essentially constant because $m$ is proportional to $r^3$ and corresponds to
the cosmological constant.
\section{Conclusion}
This work is a natural development of speculations I started to make almost 4
years ago about 4DO being an alternative formulation for relativity. At the
onset the reasoning was that if one wants to restrict 3-dimensional velocity to
the speed of light, a logical thing to do is to postulate a 4th dimension and
then state that velocity is always equal to the speed of light but can make
different angles to the 4th dimension. If then only the 3-dimensional
projection of velocity is considered this can take any value between 0 and the
speed of light. I wrote several essays elaborating on that concept which are
all available for download from the e-print archive. I made several mistakes
along the way but I don't intend to remove the respective essays because they
will allow readers to trace the track I've followed. There is one work which I
still think is important that people read \cite{Almeida02:2}, where a
comparison is established between special relativity and 4DO using the method
known as K-calculus.

The hypersphere model of the Universe is a generalization of 4DO; it is simpler
in terms of basic postulates and incorporates 4DO for everyday situations of
classical mechanics. That model is capable of explaining such puzzles as
Universe flatness or orbital velocities around galaxies as resulting entirely
from geometry, thus avoiding the discomfort of postulating enormous amounts of
dark matter. When dealing with classical mechanics problems 4DO was proven to
be equivalent to general relativity in all situations characterized by static
isotropic metrics and this equivalence is most likely extendable to all static
metric situations.

One point that made people react against 4DO in the past was the difficulty in
understanding the meaning of coordinate $x^0$. In fact geodesics of 4DO space
can be mapped to those of relativity but the same does not happen with points
in both spaces. A point where two relativistic geodesics cross is not mapped to
the crossing point of the corresponding geodesics in 4DO. A point in
relativistic space is interpreted as an event and the meaning of points in 4DO
space is difficult to grasp. It is important to consider that 4DO is a space
for optics so an elementary particle travelling in a given direction with a
known momentum should not be interpreted as a trajectory in 4DO but rather as a
plane wave that can be represented by any line normal to the wavefronts.

An example taken from optics may clarify the situation. Imagine a plane wave
travelling along the $x$ direction and another plane wave travelling at some
angle to $x$. It makes no sense to ask at what position along $x$ the two waves
meet because they meet everywhere. However, if these waves were synchronized by
some means, for instance if they were split from the same laser beam and then
redirected to converge, it would be possible to measure the length travelled by
the two waves and there would be a particular position where the two
measurements would be equal. In 4DO all trajectories are representative of
waves that were essentially all split from the same source when the big bang
happened; so even if there is a multitude of lines representing a trajectory it
is possible to define events as those points where two measurements along
different paths become equal.

In this work we took the approach of trajectories, which is the 4DO equivalent
to geometrical optics; in the future it is planned to extend this analysis with
the help of wave and Fourier optics in their 4-dimensional extensions.
  \bibliographystyle{unsrtbda}
  \bibliography{Abrev,aberrations,assistentes}   
\end{document}